\documentclass{cimento}
\usepackage{amsmath}
\usepackage{amssymb}
\usepackage{graphicx}

\title{Electromagnetic Cascades as Probes of Cosmic Magnetism}

\author{R.~{Alves Batista}\thanks{rafael.alvesbatista@physics.ox.ac.uk}}

\instlist{\inst Department of Physics - Astrophysics, University of Oxford; Denys Wilkinson Building, Keble Road, OX1 3RH, Oxford, UK}

\begin{document}

\maketitle

\begin{abstract}
The existence of intergalactic magnetic fields (IGMFs) is an open problem in cosmology and has never been unambiguously confirmed.
High-energy gamma rays emitted by blazars are unique probes of cosmic magnetism, as their interactions with pervasive radiation fields generate a short-lived charged component sensitive to intervening magnetic fields. Spatial and temporal properties of the secondary gamma rays generated in the electromagnetic cascade can provide information about the strength, power spectrum, and topology of IGMFs. To probe these fields, detailed simulations of gamma-ray propagation in the intergalactic medium are necessary. In this work the effects of magnetic fields on the spectrum and arrival directions of gamma rays are studied using three-dimensional simulations, emphasising the particular case of helical IGMFs.
\end{abstract}

\section{Introduction}

In recent years a number of TeV-emitting blazars have been observed and catalogued. Some of these objects such as 1ES 0229+200~\cite{hess2007a}, 1ES 1101-232~\cite{hess2007b}, and 1ES 0347-121~\cite{hess2007c}, present unusually hard intrinsic spectra. Various explanations for this ``problem'' have been put forth, being intergalactic magnetic fields (IGMFs) amongst the most prominent ones.

The origin of cosmic magnetic fields is an open problem in cosmology. Although seemingly ubiquitous, the existence of IGMFs has never been unambiguously confirmed (for reviews see e.g. Ref.~\cite{vallee2011a,ryu2012a}). 
There are essentially two broad classes of mechanisms for magnetogenesis: astrophysical and cosmological. The non-detection of IGMFs in cosmic voids would favour an astrophysical origin (e.g. Biermann battery), whereas its detection support a cosmological origin (e.g. phase transitions, inflation, etc). The strength and coherence length of cosmic magnetic fields can be constrained using a variety of techniques, namely Faraday rotation measurements, Zeeman splitting, imprints in the CMB, among others. None of these methods provide a lower bound on the strength of IGMFs. This can be achieved using gamma-ray-induced electromagnetic cascades in the intergalactic medium, which yields the lower bound $B \sim 10^{-17} \; \text{G}$~\cite{neronov2010a,ando2010a,chen2015a}.

A proxy for the topology of the magnetic field is the helicity, defined as $\mathcal{H} = \int_V d^3r\,\vec{A}\cdot\vec{B}$, where $\vec{B}$ is the magnetic field and $\vec{A}$ the magnetic vector potential. It has been argued~\cite{tashiro2012a,fujita2016a} that the existence of helical magnetic fields in the primeval universe could account for the matter-antimatter asymmetry without the need to introduce any beyond the Standard Model physics. In Refs.~\cite{tashiro2013a,tashiro2014a} a method was proposed to infer the helicity of IGMFs based on arrival directions of gamma rays in the $10-100 \; \text{GeV}$ energy range, finding some indications of these fields in Fermi data. The feasibility of detection of helicity in gamma-ray data has been further demonstrated in Refs.~\cite{long2015a,alvesbatista2016b}. 

With the advent of imaging air Cherenkov telescopes (IACTs), IGMFs may be studied with unprecedented precision due to the high resolution of such detectors ($\theta_{PSF} \sim 0.1^\circ$). Combined observations of blazars in the energy range $0.1-10 \; \text{GeV}$ by Fermi, and $\sim 10 \; \text{GeV} - 100 \; \text{TeV}$ by IACTs such as H.E.S.S., MAGIC, VERITAS, HAWC, and the upcoming Cherenkov Telescope Array (CTA), can provide spectral and temporal information of cascade photons. To interpret these measurements, one must understand in detail how gamma rays propagate, and consequently the interactions to which they are subject, as well as possible effects of intervening of magnetic fields upon them. 

\section{Modelling the development of the cascade}

Gamma rays emitted by distant sources can interact with ambient radiation fields such as the cosmic microwave background (CMB) and the extragalactic background light (EBL). For energies up to $E \sim 10^{14} \; \text{eV}$, the dominant interactions are pair production ($\gamma + \gamma_{bg} \rightarrow e^+ + e^-$) and inverse Compton scattering ($e^\pm + \gamma_{bg} \rightarrow e^\pm + \gamma$), where the subscript `bg' refers to the target photon field, namely the cosmic microwave background (CMB) and the extragalactic background light (EBL).

Gamma rays from distant blazars are excellent probes of the IGMFs, as their jets point approximately towards Earth. Electromagnetic cascades initiated by gamma rays from these objects are particularly interesting since intervening magnetic fields can split the electrons-positrons pairs created through pair production. These electrons and positrons can subsequently upscatter ambient photons to high energies via inverse Compton scattering. As a result of this process, the cascade is dominated by secondary gamma rays whose momenta are slightly deflected away from the source's line of sight, hence effectively increasing the angular size of the observed source, generating a halo.





For a complete Monte Carlo treatment of the effects of IGMFs on the development of electromagnetic cascades, GRPropa was developed~\cite{alvesbatista2017a}. This novel tool enables the propagation of gamma rays and secondary electrons and photons in arbitrary magnetic field configurations, accounting for all relevant energy loss processes such as synchrotron emission and adiabatic losses due to the expansion of the universe, as well as interactions with pervasive photon fields including the CMB and EBL, namely pair production and inverse Compton scattering. It is entirely based on the modular structure of the CRPropa 3 code~\cite{alvesbatista2016a}, following the implementation of Elmag~\cite{kachelriess2012a}. Details about how interactions and energy losses are implemented can be found in Ref.~\cite{alvesbatista2016b}. Distinctive features of GRPropa include weighted sampling to optimise the performance, helical stochastic magnetic fields, and transient sources, among others. 

\section{The case of 1ES 0229+200}


To illustrate the effects of magnetic fields on the gamma rays, the object 1ES 0229+200 is taken as an example. Using the computational framework mentioned in the previous section, the formation of a pair halo around the object was investigated for different magnetic fields. A stochastic Kolmogorov magnetic field grid with size $(100 \; \text{Mpc})^3$ and resolution $\sim 100 \; \text{kpc}$ was generated, and periodically repeated to cover the entire simulation volume. The coherence length was set to $\ell_c \approx 5 \; \text{Mpc}$. The EBL model by Gilmore {\it et al.}~ \cite{gilmore2012a} was adopted for all the simulations presented here.

\begin{figure}[h!]
	\centering
	\includegraphics[width=.495\columnwidth]{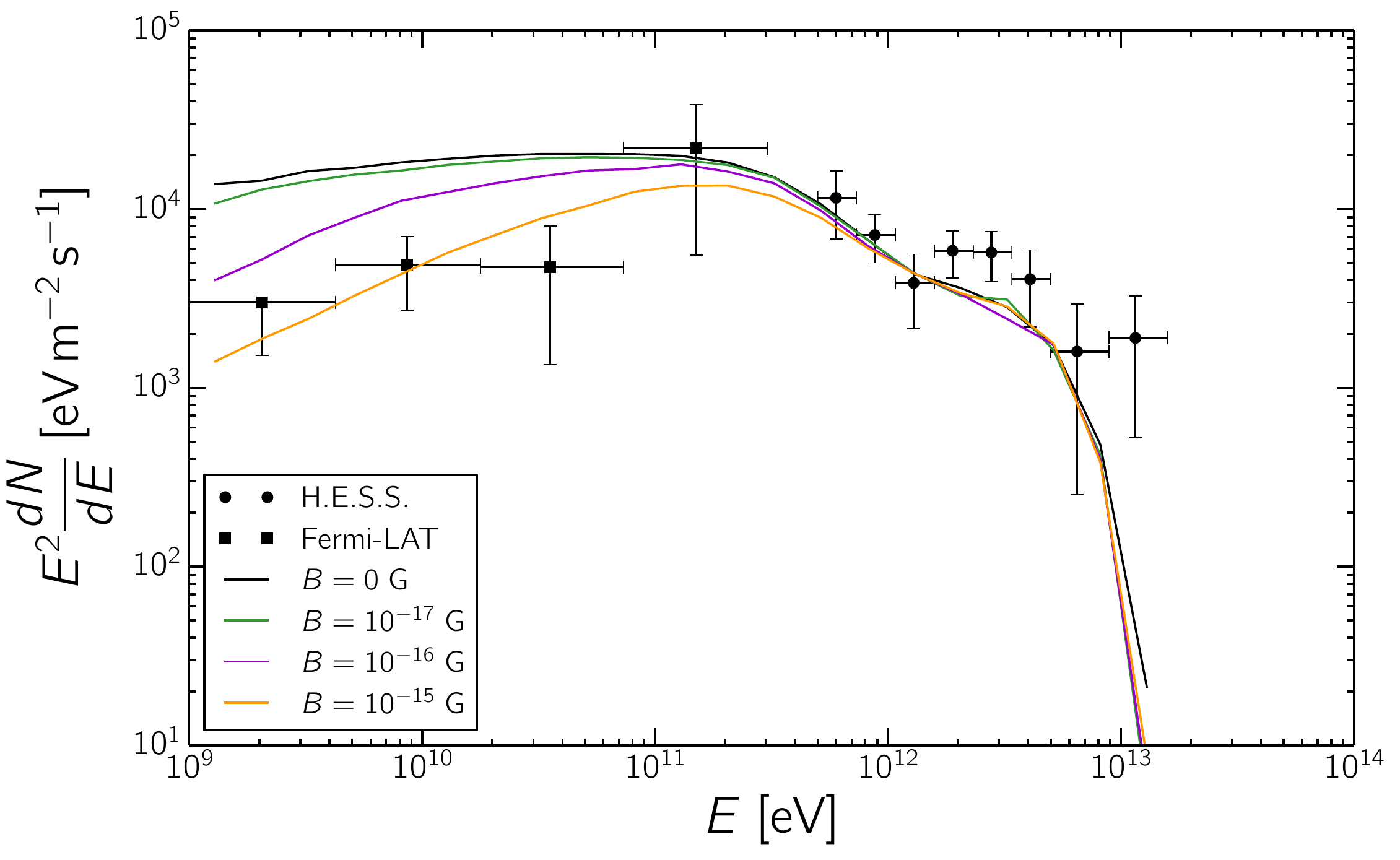}
	\caption{Energy spectrum of arriving particles from 1ES 0229+200. This object is located at $z\approx 0.14$, and is assumed to be emitting gamma rays with intrinsic spectrum $E^{-1.5}$ and cutoff energy $E_{max} = 5 \; \text{TeV}$, within a tightly collimated jet. The magnetic field has coherence length $\ell_c \approx 5 \; \text{Mpc}$ and strengths $10^{-15} \; \text{G}$ (violet), $10^{-16} \; \text{G}$ (orange), $10^{-17} \; \text{G}$ (green), and $0$ (black). Measurements by H.E.S.S. (circles) and Fermi (squares) are shown for comparison.}
	\label{fig:spec}
\end{figure}

\begin{figure}
	\centering
	\includegraphics[width=.325\columnwidth]{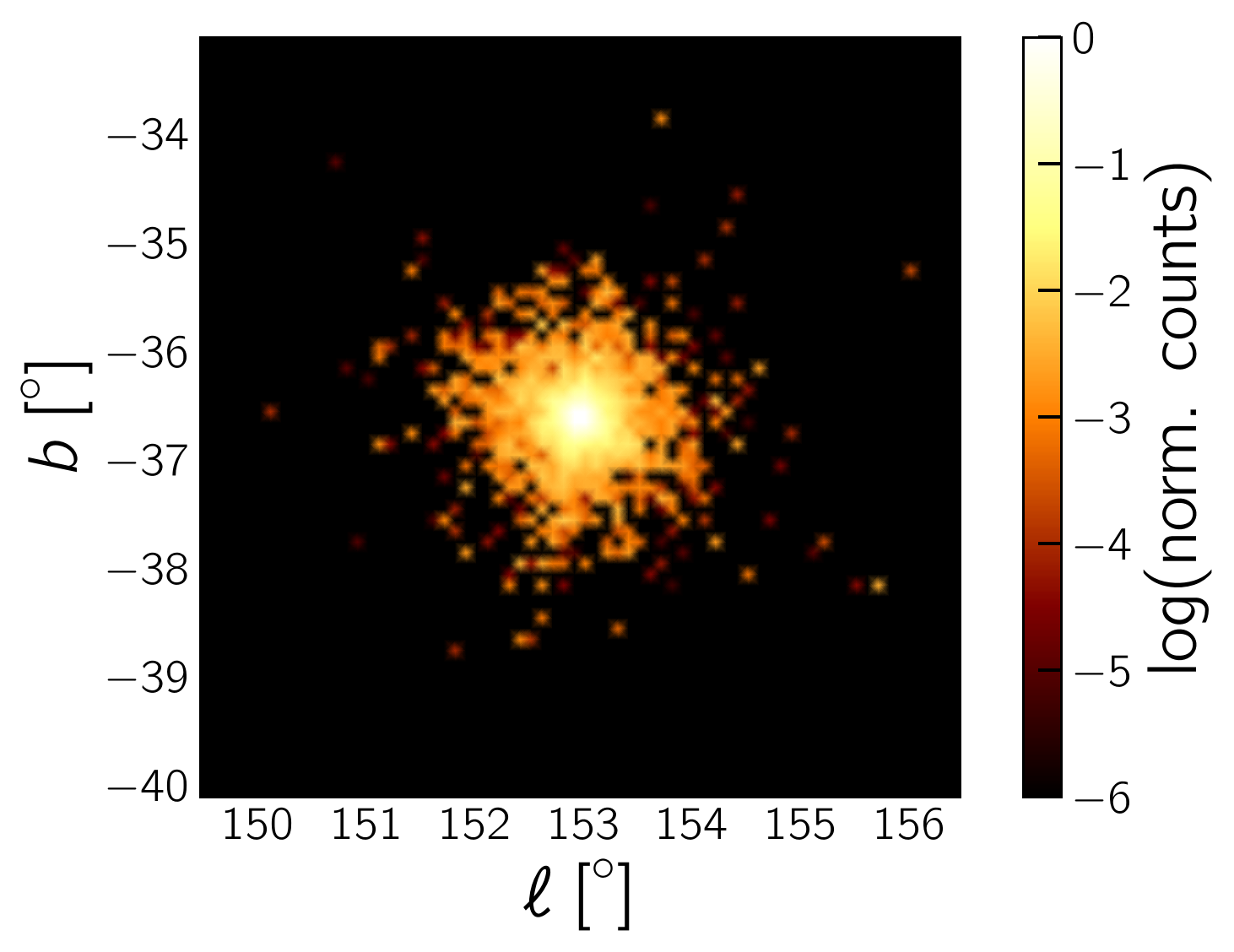}
	\includegraphics[width=.325\columnwidth]{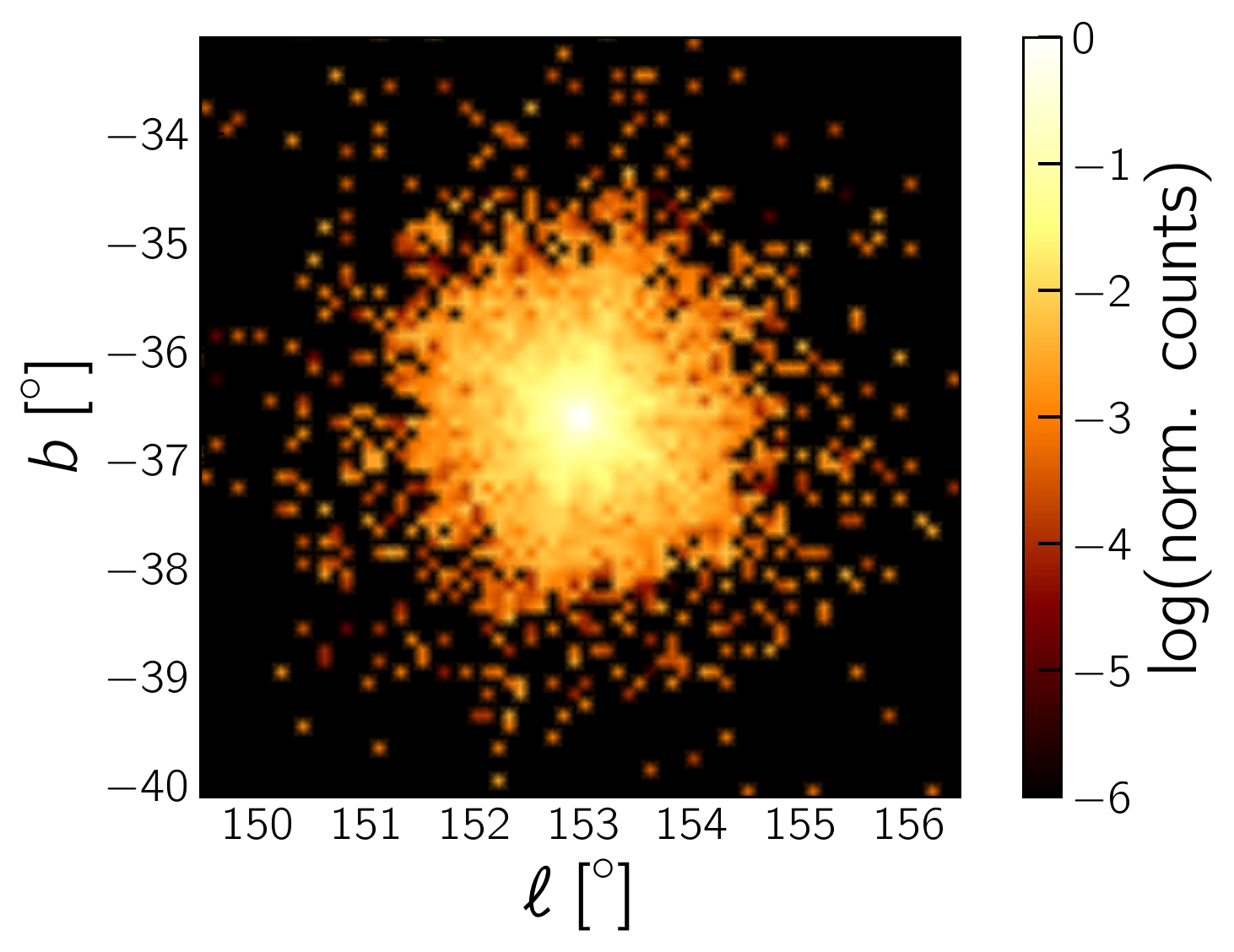}
	\includegraphics[width=.325\columnwidth]{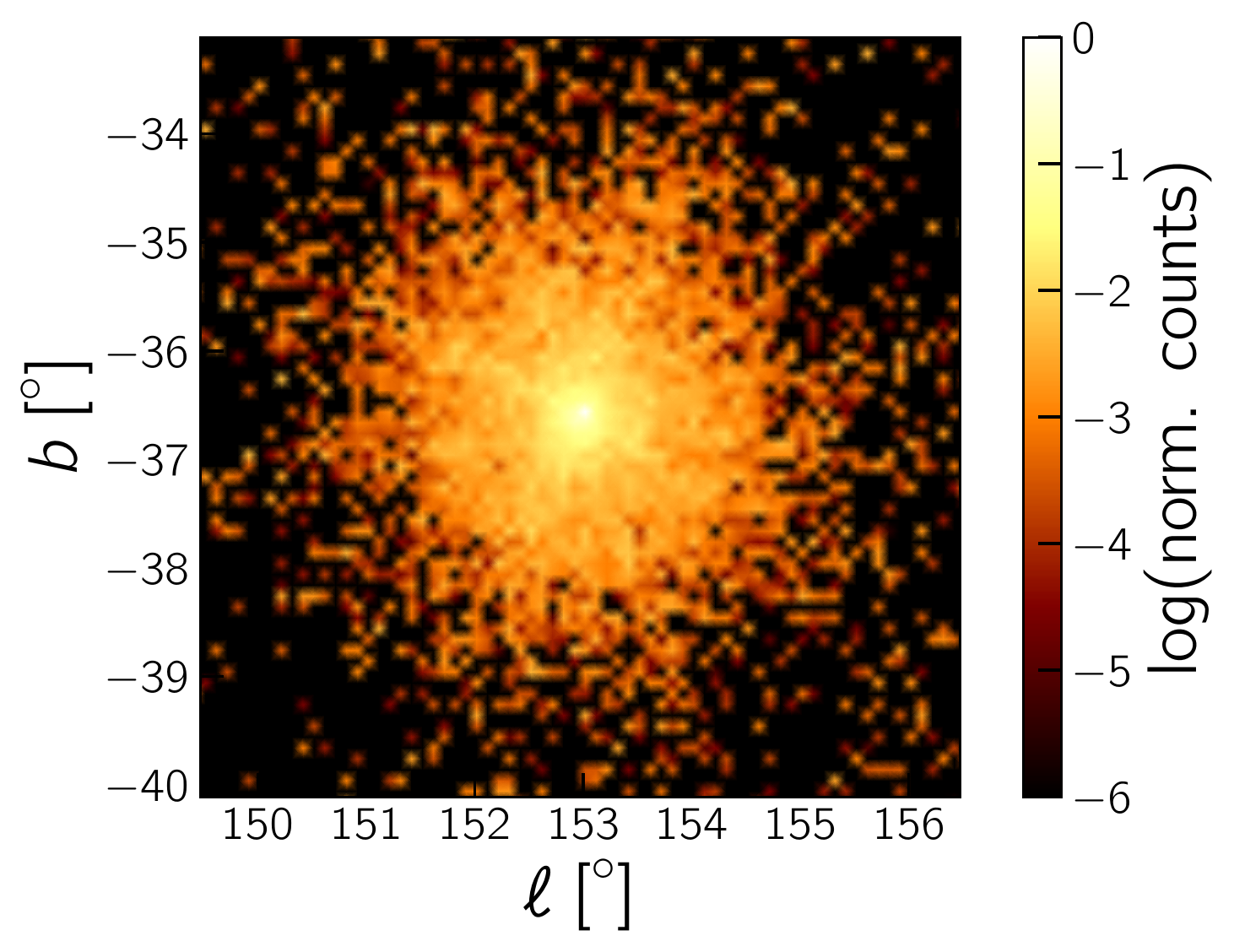}
	\caption{Simulations of the arrival directions of gamma rays emitted by 1ES 0229+200. The parameters are the same as in Fig.~\ref{fig:spec}. Each panel corresponds to a different magnetic field: $10^{-17} \; \text{G}$ (left) and $10^{-16} \; \text{G}$ (middle), and $10^{-15} \; \text{G}$ (right panel).}
	\label{fig:haloes}
\end{figure}

The simulated spectra for different magnetic field strengths are shown in Fig.~\ref{fig:spec}. The formation of pair haloes was also investigated, and the maps containing the arrival direction of the gamma rays are shown in Fig.~\ref{fig:haloes} for three magnetic field configurations: $B=10^{-15} \; \text{G}$, $B=10^{-16} \; \text{G}$, and $B=10^{-17} \; \text{G}$. 
In Fig.~\ref{fig:spec} one can see the suppression of the flux at lower energies as the magnetic field increases. This example indicates that for this specific EBL model, IGMFs with strength $B\lesssim 10^{-16} \; \text{G}$ are incompatible with measurements. The reason for this can be understood by analysing Fig.~\ref{fig:haloes}. For high magnetic fields, deflections are larger, and a considerable fraction of the gamma rays emitted by the source will skim past or completely miss Earth.

\section{Helical magnetic fields}

\begin{figure}[h!]
	\centering
	\includegraphics[width=.30\columnwidth]{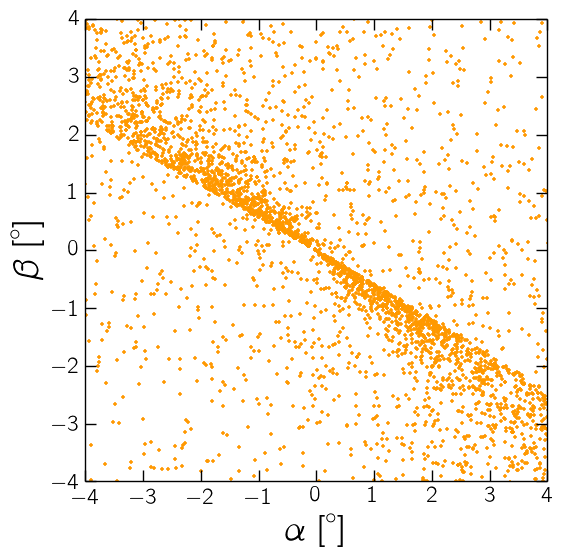}
	\includegraphics[width=.30\columnwidth]{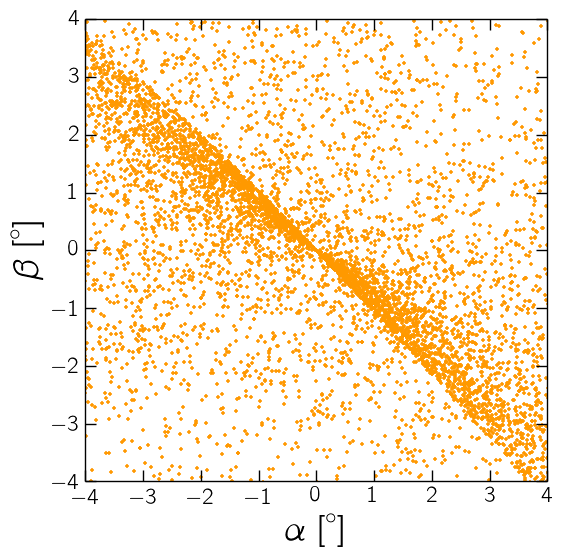}
	\caption{Distribution of arrival directions of the gamma rays. The angles $\alpha$ and $\beta$ refer to an arbitrary coordinate system. The parameters are the same as in Fig.~\ref{fig:spec}, except for the energy, which was set to $E = 30 \; \text{TeV}$. The magnetic field used here is $B=10^{-15} \; \text{G}$, with a coherence length of $\ell_c \approx 180 \; \text{Mpc}$. Each panel corresponds to a different helicity: maximally negative (left), and maximally positive (right panel).}
	\label{fig:helicity}
\end{figure}

Using simulations obtained with GRPropa the formation of blazar pair haloes in the presence of intervening helical magnetic fields was studied. Once more, the object 1ES 0229+200 was chosen, assuming $\ell_c \approx 180 \; \text{Mpc}$ and $B = 10^{-15} \; \text{G}$. The gamma rays are emitted with energies $E=5 \; \text{TeV}$ within a collimated jet pointing straight at Earth. The large coherence length used here is compatible with expectations from a number of magnetohydrodynamical simulations~\cite{jedamzik2011a,saveliev2013a}, which suggest that helical magnetic fields present a Batchelor spectrum with large coherence lengths ($\ell_c \gtrsim 10 \; \text{Mpc}$). This value also results in a pattern clearly visible in sky maps, as shown in Fig.~\ref{fig:helicity} for two different helicity assumptions: maximally positive and maximally negative. 

From Fig.~\ref{fig:helicity} it is clear that the handedness of the observed pattern is related to the helicity and present opposite behaviours for opposite helicities. The pronounced filamentary feature arises from the fact that for large coherence lengths the field is effectively uniform. The component perpendicular to the jet in the immediate vicinities of the blazar determines the direction of deflection of the bulk of gamma rays.

The effects of helical fields on the formation of blazar pair haloes have been studied in Refs.~\cite{long2015a,alvesbatista2016b} for the case of single sources. Using methods such as the Q-statistics~\cite{tashiro2013a} or S-statistics~\cite{alvesbatista2016b}, one can estimate the helicity of the intervening field. Nevertheless, a proper assessment of the helicity requires the analysis of the arrival directions of gamma rays from many sources.

\section{Outlook}

A few examples of the effects of magnetic fields on cascades were shown. The spectrum is suppressed at energies $\lesssim 100 \; \text{GeV}$ proportionally to the magnetic field, as shown in Fig.~\ref{fig:spec}. This confirms the lower bounds derived by a number of authors~\cite{neronov2010a,tavecchio2010a,vovk2012a}. One should, however, be cautious when drawing such conclusions. The naive expectation that the flux is suppressed due to the production of pairs that are deflected away from Earth, or delayed if the source is transient, is not completely accurate. Deflections of pairs whose initial momenta do not point to Earth can contribute to the production of secondary photons which reach the observer. Therefore, the contribution of blazars with misaligned jets should also be taken into account, particularly when estimating the gamma-ray background at $\sim \; \text{GeV}$ energies. 

To date, most of the literature studying the formation of pair haloes has addressed the extension of haloes, but not their morphology. The magnetic helicity relates to the latter, and is a proxy for processes taking place in the primeval universe. The handedness of the helicity gives indications of the mechanisms whereby magnetic fields originated; for electroweak baryogenesis, for instance, one expects left-handed magnetic fields. Therefore, there is a clear motivation for three-dimensional Monte Carlo simulations, as opposed to one-dimensional setups or semi-analytical approaches.

Enhancements of current experiments and new ground-based observatories such as CTA will deepen our knowledge about cosmic magnetic fields. In the next decade we will likely witness a leap forward in this field, with refined limits on the strength of IGMFs based on spectral measurements as well as imaging techniques. New methods should be devised to relate the topology of IGMFs with the morphology of the observed gamma rays. All these measurements, combined, may potentially have profound implications for fundamental physics and early-universe cosmology.

\acknowledgments

I am indebted to Andrey Saveliev for his inumerous contributions to the development of the code and for comments about this manuscript.

\bibliographystyle{varenna}

\end{document}